\documentclass[twocolumn,showpacs,aps,prb]{revtex4}

\usepackage{graphicx} 
\usepackage{dcolumn} 
\usepackage{bm} 

\begin{document}
\preprint{Physical Review B}

\title{Transport through the intertube links between two parallel carbon nanotubes }

\author{Jia-Lin Zhu}
\email[Electronic address: ]{zjl-dmp@tsinghua.edu.cn}
\affiliation{%
Department of Physics, Tsinghua University, Beijing 100084,
People's Republic of China}
\author{Fu-Fang Xu}
\affiliation{%
Department of Physics, Tsinghua University, Beijing 100084,
People's Republic of China}
\author{Yu-Feng Jia}
\affiliation{%
Department of Physics, Tsinghua University, Beijing 100084,
People's Republic of China}

\date{\today}

\begin{abstract}
Quantum transport through the junction between two metallic carbon
nanotubes connected by intertube links has been studied within the
TB method and Landauer formula. It is found that the conductance
oscillates with both of the coupling strength and length. The
corresponding local density of states (LDOS) is clearly shown and
can be used to explain the reason why there are such kinds of
oscillations of the conductances, which should be noted in the
design of nanotube-based devices.

\end{abstract}
\pacs{73.22.-f, 73.23.-b, 73.40.-c}

\maketitle
\section{Introduction}

Since the discovery of single-walled carbon nanotube (SWNT) by
Iijima,\cite{Iijima} it had inspired great interests during the
past years, due to its unique structural and electronic qualities
\cite{Dresselhaus} and great potential applications for
nanodevices. Many prototypes of electronic devices based on carbon
nanotubes have been realized in experiments, such as field-effect
transistors,\cite{device1} single-electron
transistors,\cite{device2} rectifiers,\cite{device3} nonvolatile
random access memory\cite{device4} and so on. SWNT can be metallic
or semiconducting depending on its diameter and helicity. Because
of the small size in diameter, transport through carbon nanotubes
is in ballistic regime. A perfect metallic SWNT has two crossing
energy bands near the Fermi level ($E_F$), namely, the
$\pi$-bonding and $\pi$-antibonding bands. These two bands act as
conducting channels and  contribute two conductance quanta to the
low bias conductance $G(=2G_0=4e^2 /h)$ according to the Landauer
formula.\cite{Datta}

Many efforts have been made to manipulate the electronic
properties of carbon nanotubes, such as doping, deforming
\cite{Yu} ,functionalization,\cite{Strano} and construction of
various structures consisting of nanotubes. Interface or junction
becomes a critical issue for mesoscopic transport in many cases.
In the previous researches, carbon nanotubes have demonstrated
quite different properties, such as ballistic
transport\cite{ballistic}, Coulomb blockade\cite{qd} and Luttinger
liquid behaviors\cite{ll}. It is the contact effects and the inner
scattering mechanisms that play the key role and make the
nanotubes fall into different regimes. The influence of interface
between SWNT and metal electrode on the conductance has been
studied both in the frame of TB model\cite{Xue} and {\it ab
initio} calculation.\cite{Taylor} These results showed that the
interface affected the transport strongly.

The junctions of carbon nanotubes, such as cross, Y and T shapes,
have been studied experimentally and theoretically, and showed
unusual physical properties. Two carbon nanotubes with different
diameters can be linked by pentagon and heptagon pairs in the
junction.\cite{Dresselhaus} The stable junctions of various
geometries have been fabricated with the help of transmission
electron microscope in experiments.\cite{Terrones} Molecular
dynamics calculations were implemented to simulate the bombardment
of nanotubes and demonstrated that the crossed nanotubes could be
welded.\cite{Krasheninnikov} Under pressure, the intertube links
are formed between the zigzag nanotubes in the SWNT ropes.
Meanwhile, it was pointed that similar interlinking C-C bonds do
not form between the (6, 6) parallel tubes even if they were
deformed under a very high pressure.\cite{Yildirim}

The study of the transport through these nanotube junctions is an
important subject. It have been showed that intertube conductance
through the junctions of two crossed single-walled carbon
nanotubes was strongly dependent on the atomic registries between
the two tubes.\cite{Fuhrer} A significant conductance is allowed
through the junction under relatively weak contact forces. These
interlinking bonds survive even after the contact forces are
released and the whole structure is fully relaxed.\cite{Buldum} As
for two parallel SWNTs, it is expected that they are linked by the
van der Waals interaction when their distance is larger than
$D_{vdW}=s_{vdW}+2R$,($s_{vdW} \sim$ 3.3 \AA, R is the radius of
the nanotube). By constraining them with the distance $D<D_{vdW}$,
the interlinking bonds are easily formed between two tubes. The
results of calculation indicated that the junction provided good
transport in some atomic registries.\cite{Dag}

How the coupling strength and length affect the transport through
the intertube links between two parallel nanotubes? This is an
interesting and valuable question because it is one of requisite
steps to realize the pure-nanotube based devices. However, these
effects have not been investigated systematically so far. Here the
TB method is taken to deal with the junction of the parallel
nanotubes and these effects are explored within the Landauer
formula. The configuration is taken as that the coupling part of
the two parallel metallic zigzag nanotubes is placed symmetrically
and two atoms are connected in each unit cell (u.c.).

\section{Model and Formula}
The $\pi$-electron tight-binding (TB) model and the Green's
function method have been used to accomplish the calculation. It
is proved to be valid for states near the $E_F$ by previous {\it
ab initio} calculations.\cite{Rubio} According to the band
structure calculations, the states crossing the $E_F$ are
predominantly $\pi$ like in character. The electronic and
transport properties of carbon nanotubes should be dominated by
$\pi$ electrons. It can provide satisfied results to use the
nearest-neighboring tight-binding model with one $\pi$ electron
per carbon atom.\cite{song} In our calculations, on-site energies
are set to zero and all the hopping parameters between nearest
sites equal to 2.66 eV.\cite{Chico}

In our model, we take the configuration containing two parallel
open ended SWNTs. In figure 1(a), the geometric scheme of the
junction is given. The two nanotubes are linked by the bonds
formed between two nearest C atoms in each unit cell. Figure 1(b)
gives the view from one end along the axis. We can see that two
nearest atoms are connected by the intertube link, which is both
mechanical and electronic. Once the coupling part is included, the
length of nanotubes chosen as conductor has no effect on the
results for certain numbers of coupling unit cells. We only take
the coupling part as conductor in our calculations, and take the
nanotubes beyond the conductor part as leads, as shown in figure
1(c). Each nanotube is kept regular shape because the deformation
is small and the effects of deformation on transport can be
neglected. As mentioned earlier, we take $\pi$-electron
tight-binding model to deal with the junction, which contributes
the main part of conductance. The link is covalent $\sigma$ like
bond formed between two opposite $\pi$ electrons and the coupling
strength $t_c$ is taken to describe the link between them.

\begin{figure}
\includegraphics[angle=0,width=0.39\textwidth]{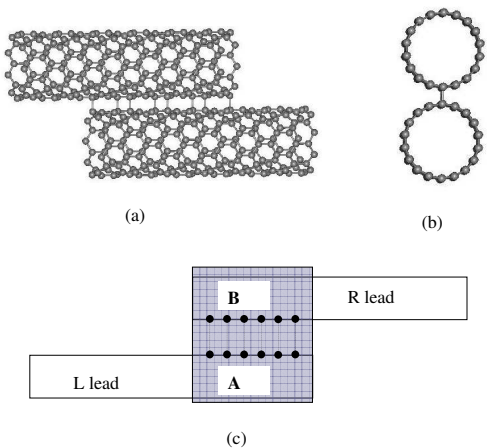}
\caption{\label{FIG:model}Schematic geometry of the junction. (a)
Two nanotubes connected by intertube link. (b) View along the axis
from one end. (c) The structure is divided as L(left) lead,
conductor and R(right) lead to implement the calculation.}
\end{figure}

It should be mentioned, as some theoretical researches pointed,
isolated primary metallic SWNTs except the armchair SWNTs generate
a pseudogap near the $E_F$ due to the curvature. (6,0) carbon
nanotube is always metallic, the other (3I,0) zigzag nanobubes (I
is a integer), have a minigap which is on the order of $10^{-1}$
eV, and the magnitude is depend inversely on the square of tube
radius \cite{gulseren,zolyomi,ouyang}. The gap is so small that
the bias voltage which need not very high can strike the gap to
realize transport. It can be smeared by the small thermal
fluctuation at finite temperature also. As many previous
researches, here we do not consider the curvature effect in the TB
model either.

Within the Landauer formalism, the conductance $G(E)$ of the
system can be calculated as a function of the energy E of incident
channels from the lead:

\begin{eqnarray}\label{EQ:T}
G(E)=G_0Tr[{\Gamma_R}{G^r_c}\Gamma_L{G^a_c}]
\end{eqnarray}

where $\Gamma_{L(R)}$ is the coupling matrix between the left
(right) lead and the conductor, and $G^r_c$ is the retarded
Green's function which can be written as

\begin{eqnarray}\label{EQ:Gc}
G^r_c=(E-H_c-\Sigma_L-\Sigma_R)^{-1}, G^a_c=G^{r+}_c
\end{eqnarray}

Here $H_c$ is the Hamiltonian matrix of the conductor that
represents the interaction between the atoms in the conductor, and
$\Sigma_{L(R)}$ is the self-energy function that describes the
effect of the left (right) lead. Once $\Sigma_{L(R)} $ is known,
$\Gamma_{L(R)}$ is easily obtained as

\begin{eqnarray}
\Gamma_L=i(\Sigma_L-\Sigma^+_L),   \Gamma_R=i(\Sigma_R-\Sigma^+_R)
\end{eqnarray}

$\Sigma_L(R)$ can be computed along the lines of Ref. 30 by using
the surface Green¡¯s function matching theory.
$\Sigma_L=H^+_{LC}\textbf{\emph{g}}_L{H_{LC}},
\Sigma_R=H_{CR}\textbf{\emph{g}}_R{H^+_{CR}}$.
$\textbf{\emph{g}}_{L(R)}$ is Green's function for semi-infinite
lead, and can be calculated through transfer matrices T and
$\bar{T}$.
\begin{eqnarray}
\textbf{\emph{g}}_L=({E-H_{00}-H^+_{01}{\bar{T}}})^{-1},
\textbf{\emph{g}}_R=({E-H_{00}-H_{01}T})^{-1}
\end{eqnarray}

Transfer matrices T and $\bar {T}$  can be easily computed via an
iterative procedure.

Once $G^r_c$ are known, the total density of states (DOS) can be
easily computed,

\begin{eqnarray}\label{EQ:rdos}
\rho=-Im[Tr({G^r_c})]/\pi
\end{eqnarray}

We can calculate the Local density of states (LDOS) for the
$j_{th}$ site  in the conductor from the diagonal elements of
$G^r_c$,
\begin{eqnarray}\label{EQ:Ldos}
\rho_j=-Im[(G^r_c)_{j,j}]/\pi
\end{eqnarray}

In the case of transport through the interface between A and B,
$H_c$ and $G^R_c$ can be writen
\begin{displaymath}
H_c = \left( \begin{array}{ccc}
H_{AA}& H_{AB}  \\
H_{BA}& H_{BB} \end{array} \right), G^r_c= \left(
\begin{array}{ccc}
G_{AA}& G_{AB}  \\
G_{BA}& G_{BB} \end{array} \right)
\end{displaymath}

and the conductance can be written as\cite{Nardelli}
\begin{eqnarray}
G(E)=G_0Tr[{\Gamma_A}{G^r_{AB}}\Gamma_B{G^a_{BA}}].
\end{eqnarray}

\section{Results and Discussion}

\begin{figure}
\includegraphics[angle=0,width=0.49\textwidth]{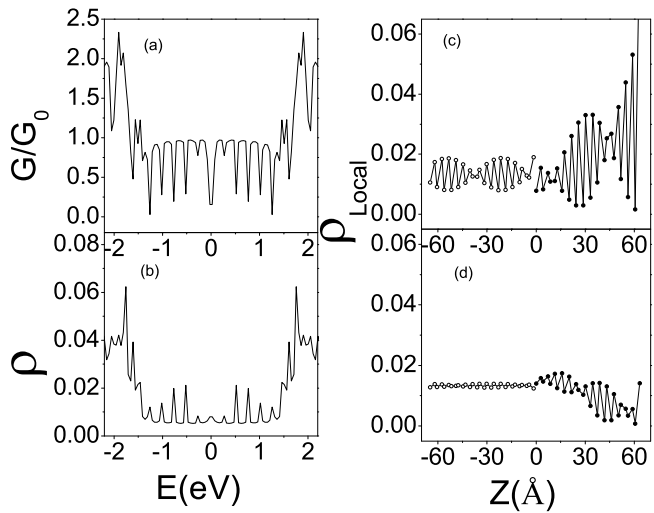}
\caption{\label{FIG:GDOS}Quantum conductance G (a) and the
corresponding total density of states (b) as a function of energy
E for (9,0)-(9,0) SWNT with 15 coupling u.c. The local density of
states of the sites in the coupling line for E=-0.272eV,
G=0.77$G_0$ (c) and E=-0.223eV, G=0.97$G_0$ (d) are plotted.}
\end{figure}

The links of C atoms between the two parallel nanotubes provide a
channel for the transport, but its value is limited to not above
the unit of quantum conductance near the $E_F$ due to the
scattering mechanism. Electrons in the two nanotubes interfere
through the interactions between them, and become extended or
localized in these links. The transport through the junction can
be tuned by the coupling strength and length. In order to explore
the inner mechanism, we have calculated the LDOS of the sites in
the coupling line.

Fig 2(a) and 2(b) show the quantum conductance and the
corresponding DOS for 15 coupling unit cells respectively. There
are many peaks in the DOS, and the corresponding valleys of
conductance appear at the same energy. These peaks in the DOS are
local peaks which give rise to the valleys of conductance
function. With longer coupling length, the peaks in the DOS and
valleys in conductance are much more complicated. The conductance
at certain energy or the average value near $E_F$ experiences
oscillations with the coupling length increasing, which will be
illustrated later. Fig 2(c) is the LDOS at E=-0.273eV and the
corresponding $G=0.77G_0$, which is a valley in the conductance
function, and we find that the density fluctuations are strong. On
the contrary, the LDOS at E=-0.223eV has weak fluctuations in the
conductor part, and the corresponding conductance reaches a high
value, $G=0.97G_0$.

\begin{figure}
\includegraphics[angle=0,width=0.45\textwidth] {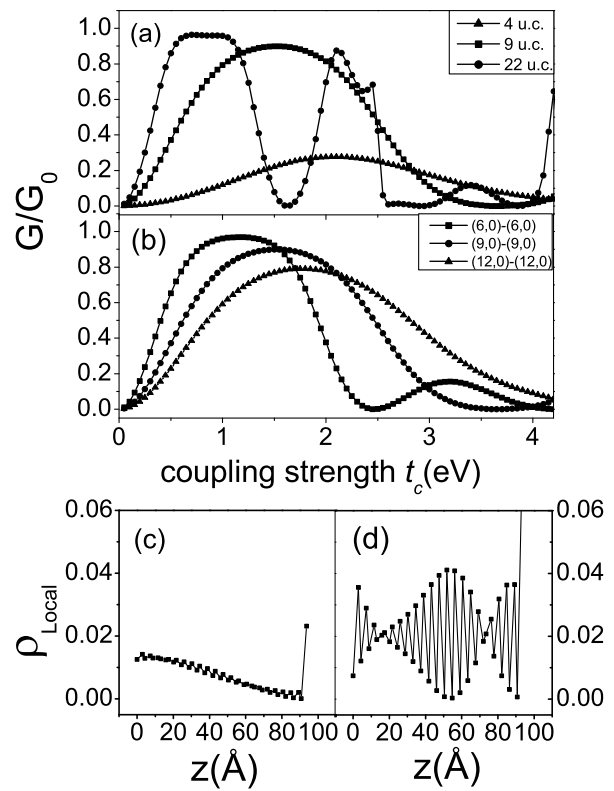}
\caption{\label{FIG:3}(a) Conductance at E=-0.13eV versus coupling
strength for (9,0)-(9,0) SWNT with 4 (triangle), 9 (square) and 22
(circle) coupling u.c. respectively. (b) Conductance versus
coupling strength for the configuration of (6,0)-(6,0)(triangle),
(9,0)-(9,0) (square) and (12,0)-(12,0) (circle) SWNT with 9
coupling u.c respectively. (c) The LDOS of sites in coupling line
of the nanotubes, with coupling strength $t_c=0.8 eV$. (d) The
LDOS with coupling strength $t_c=1.6 eV$.}
\end{figure}

Considering the pseudogap in the zigzag carbon nanotube due to its
curvature, which is not included in our model, we take the
incident energy near the $E_F$ but outside the small gap. We
investigate the dependence of the conductance on the coupling
strength at E=-0.13eV for two (9,0) metallic zigzag nanotubes. The
conductance undergoes damped oscillations, and reaches almost zero
after a few oscillations as illustrated in figure 3(a) for
different coupling strength. This oscillating phenomena is very
similar with the recent results in the experiment for the squashed
carbon nanotubes,\cite{Navarro} where the conductance oscillations
are caused by the open-close cycles of gap due to deformation.
Here the oscillations have different origin, and are caused by
delocalization and localization due to the effective deformation
induced by intertube link. The effective deformation does not
produce the gap, but causes localization and delocalization, and
then affects the transport channel between the two nanotubes. In
figure 3(a), the conductance is plotted as the function of
coupling strength for 9, 14 and 22 u.c. respectively. We can see
that the conductance is more sensitive to the coupling strength
and experiences more oscillations before getting blocked if the
coupling length is longer. In figure 3(b), the conduction versus
coupling strength is given for three different configurations. If
the coupling length is fixed, it needs higher coupling strength to
reach the first maximum of conduction for the two nanotubes with
bigger radii and the corresponding maximum value is smaller.
Figure 3(c) and 3(d) give the LDOS of the sites in the coupling
line for $t_c=0.8 eV$ and $t_c=1.6 eV$ and the corresponding
conductance are 0.96$G_0$ and 0.003$G_0$ respectively. We can see
from the figure that the LDOS fluctuations are weak when the
conductance is high.

\begin{figure}
\includegraphics[angle=0,width=0.49\textwidth]{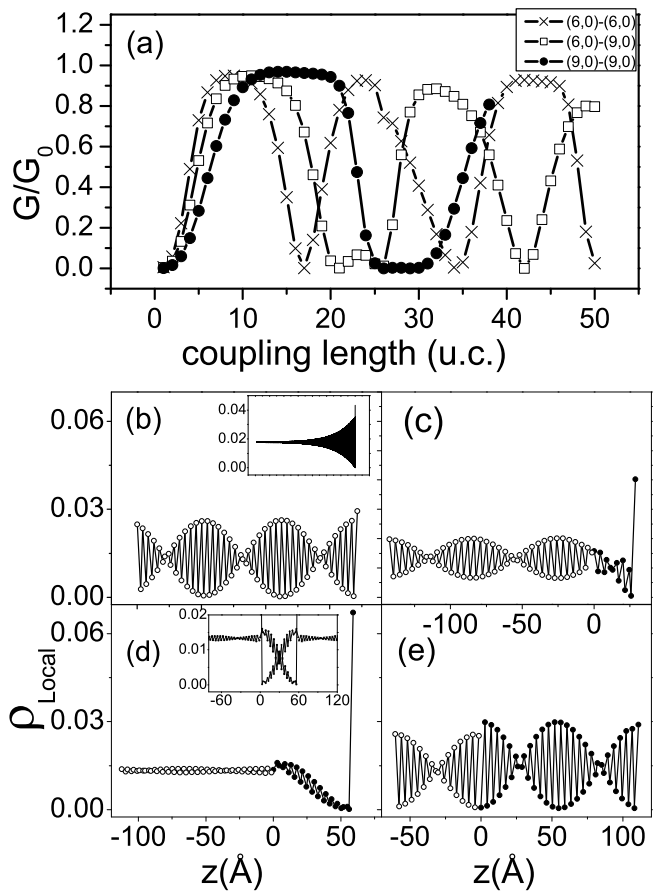}
\caption{\label{FIG:LDOS}(a) Conductance at E=-0.13eV versus
coupling length for (6,0)-(6,0), (6,0)-(9,0) and (9,0)-(9,0)
parallel nanotubes respectively. (b)-(e) The LDOS of sites in the
coupling line of nanotubes, without coupling, 7, 14 and 27
coupling u.c respectively. The inset of (b) is plotted for a
semi-infinite SWNT. The inset of (d) is the LDOS for the whole
junction including both left part and right part. We have plotted
the LDOS of sites beyond the conductor (hollow circle) for
comparison. }
\end{figure}

The coupling length has significant effects on the conductance. We
take a weak coupling value $t_c=1.4 eV$ to investigate the quantum
conductance through the intertube link. As indicated in Figure
4(a), the conductance at E=-0.13eV does not change monotonically
with the increasing of coupling length, but shows an oscillating
behavior. Starting from the configuration that only one unit cell
is connected, the conductance increases with the numbers of
coupling unit cells increasing. After reaching a peak value, it
decreases and gets a minimum value which is almost zero at certain
number of coupling unit cells. When we increase coupling length
further, it increases again. The diameters of the two nanotubes
are related to the period of oscillations. It needs more coupling
length to complete one oscillation for the nanotubes with bigger
radii, as illustrated in Figure 4(a). If the coupling length is
fixed, the conductance is higher for the nanotubes with bigger
radii before reaching the first maximum. The LDOS is a constant at
each site for a perfect infinite single-walled carbon nanotube.
But for the semi-infinite nanotubes, as shown in figure 4(b), the
LDOS at two neighboring sites has a quick oscillation and the
envelope has a long period oscillation. The magnitude of envelope
oscillation decreases from the end and becomes almost constant far
away from the end, as shown in the inset of fig 4(b).

Figure 4(c)-(e) give three typical results of the LDOS in the
coupling line. To compare them with each other, we have plotted
the LDOS of the same length. The hollow circles are LDOS of the
sites at the left lead, and their coordinates are negative. We
find that the conductance oscillation behavior is induced by
localization and delocalization, which is common in
low-dimensional systems, though the mechanism in the current
structure is more complicated. The interaction between the
electrons in the two nanotubes takes effect through the intertube
link. It brings the effective deformation, thus causes the
localization of electrons. When high conductance is gotten, the
localization disappears and the LDOS have the weakest
fluctuations, as in figure 4(d). The LDOS fluctuations in the
leads are also weak once the conductance is high. On the contrary,
the LDOS have strong fluctuations and the electrons are localized
in the conductor region, as shown in figure 4(e).

For 14 u.c. coupling 4(d), the conductance has a high value
G=0.97$G_0$ since the LDOS has very weak fluctuations. The
electrons can transport through the junction easily. The inset of
figure 4(d) shows the LDOS of the whole structure including the
two nanotubes, electrons have almost entirely transport to the
other nanotube at the end. When 26 u.c. was coupled 4(e), the
conductance is 0.002$G_0$. The electrons are localized  in the
conductor region and hardly transport through the intertube link,
thus the LDOS form strong oscillations in the conductor. Figure
4(c) is the LDOS for 7 coupling u.c. and the corresponding
conductance is 0.56$G_0$. The magnitude of conductance can be
estimated from the LDOS oscillation. The conductance is higher
when the oscillations are weaker.

We have also calculated the quantum transport through an infinite
metallic zigzag nanotube with an additional finite nanotube
nearby, which is placed parallel to the infinite one. The results
show that the quantum conductance oscillates too, but within $G_0$
and 2 $G_0$. The additional part destroys one of the two channels,
and the other one is kept unaffected.

\section{Summary}

We have studied the transport through intertube links between two
parallel metallic zigzag SWNTs, using Green's function method and
Landauer formula. It is found that the conductance shows
oscillating behaviors. Damped oscillations arise as the coupling
strength increases, and the transport gets blocked after a few
oscillations. Quantum conductance through the junction also shows
oscillations when increasing coupling length. The conductance
reaches the maximum which is near 1$G_0$ at some coupling length
and gets almost zero at some other length. Longer coupling length
is needed to finish one oscillation as the radius of nanotube
increases. The LDOS has been investigated and the results indicate
that the conductance has relation to the amplitudes of the
fluctuations of LDOS. When the electrons can transport through the
junction with a high (low) conductance, the LDOS shows weak
(strong) fluctuations. The electrons in the two semi-infinite
nanotubes interfere through the intertube link. These links cause
the effective deformation of the nanotubes, which induces the
localization and delocalization of the electrons in the conductor.
Consequently, the LDOS at the sites of coupling line has weak or
strong fluctuations.

A detailed description of the quantum transport through the
junction of two parallel zigzag nanotubes is given and further
research is needed to achieve a comprehensive understanding in
order to construct the pure nanotube-based devices. The work
suggests that it should be very careful when constructing devices
in nanostructures since the interface or junction is very
important for SWNT-based structures. The results implicate that
this structure may be utilized to design nanodevices since the
intertube transport can be tuned by changing the strength or the
length of coupling.

\begin{acknowledgments}
Financial support from NSF-China (Grant No. 10374057 and 10574077)
and ``973'' Programme of China (No. 2005CB623606) is gratefully
acknowledged.
\end{acknowledgments}


\end{document}